\documentclass[12pt]{article}

\usepackage{graphicx}
\usepackage{epsfig,amsmath}
\usepackage{hyperref}
\normalbaselines

\begin{document}
\title{The Blast Energy Efficiency of GRBs}

\author{David Eichler\footnote{Physics Department,
Ben-Gurion University, Beer-Sheva 84105, Israel;
eichler@bgumail.bgu.ac.il} \;and Daniel
Jontof-Hutter\footnote{Physics Dept., Ben-Gurion University,
Beer-Sheva 84105, Israel; djontofhutter@hotmail.com}} \maketitle

\begin{abstract}
Using data mostly assembled by previous authors, we consider the
linear correlation between the apparent radiative efficiency
$\epsilon_{\gamma}$ (defined as the ratio of isotropic equivalent
radiative output to inferred isotropic equivalent kinetic energy
of the blast) and $E_{peak}^{\alpha}$ where $1.4<\alpha<2$, for 17
of 22 GRBs (Lloyd-Ronning and Zhang, 2004). We note in a
quantitative manner that this is consistent with the hypothesis
that $\epsilon_{\gamma}$ and $E_{peak}$ are influenced by viewing
angle. We suggest a more general theoretically derived expression
for this correlation that could be tested with a richer data set.
If the reduction in both $\epsilon_{\gamma}$ and $E_{peak}$ is due
to viewing angle effects, then the actual radiative efficiency is
$\sim 7$. We also find preliminary evidence (with a small sample)
for a separate class of weak GRB afterglows.
\end{abstract}

\section{Introduction}
It is well understood that the highly super-Eddington luminosities
associated with GRB are liable to put most of their energy into a
baryonic wind if the energy release is in a position to drag
matter outward. One solution for this (Meszaros and Rees 1994) is
that at distances $\ge 10^{13}$cm from the central burster,
internal shocks in a baryonic outflow  release some fraction of
the bulk expansion energy by accelerating particles, which then
radiate $\gamma$-rays. This probably predicts that typically 10 to
50 percent of the energy can be recovered and put back into
radiation. It also predicts that the $\gamma$-rays are always
accompanied by baryonic outflow along the same direction to within
$1/\Gamma$.

    Alternatively, it may be supposed that  the energy release is
originally   devoid of baryons (e.g. if the energy emerged along
event horizon -threading field lines [Eichler and Levinson, 1993])
and the the baryon content of the GRB fireball is whatever it
swept up subsequently, either from the sides (Eichler and Levinson
1999, Levinson and Eichler 2003), or from the ambient material
into which the fireball expands (e.g. Meszaros and Rees 1992,
Lyutikov and Blandford 2004). If the fireball were able to sweep
up ambient material without having any baryons originally, then
afterglow would be "guaranteed" provided that  the ambient medium
has sufficient density. However, if the asymptotic Lorentz factor
$\Gamma_{a}$ of the fireball were too large, then it would not
pick up ambient matter; early baryon loading or an initial baryon
content, which  keeps
 $\Gamma_{a}$ from getting too large, is probably necessary for the GRB to have afterglow.
  Moreover, the
 Poynting flux may be considerably less than the $\gamma$-ray flux
 from the central object, and baryon loading from the side
 downstream
 of such a fireball's point of origin
  could reclaim some of
 the $\gamma$-ray energy for generating afterglow
 if its optical depth exceeded unity. The case can thus be made that
 baryon loading from the sides of such a fireball, as it exits a baryon
 rich surrounding, can enhance the blast efficiency until it is
 of order unity. (In this paper, blast efficiency, $\epsilon_{k}$ refers to the
 isotropic equivalent kinetic energy $E_k$ divided by the isotropic
 equivalent
 $\gamma$-ray energy output, $E_{\gamma,iso}$. The $\gamma$-ray efficiency, $\epsilon_{\gamma}$, refers to the
 inverse of the blast efficiency, and either quantity can be greater than
 unity.)

 There is no guarantee that early baryon loading from the sides penetrates the entire
 fireball; there exists the logical possibility that one could
 have a $\gamma$-ray bright GRB with little or very weak
 afterglow. Previous estimates
 (Levinson and Eichler 2003, Eichler and Levinson 2004 [EL04])
 suggested a picture in which the penetration is only 1/3 to 1/2
 of the angular distance to the center from the outside, depending on the duration.
 Longer bursts allow greater penetration because  the penetrating
 baryons are freely streaming neutrons, and the larger the radius
 they are allowed to penetrate, the lower the Lorentz factor of
 the fluid at the surface from which free streaming begins.
 Observers
 close to the axis of symmetry might therefore see afterglow only
 if the spread in the afterglow beam, which is smeared by an
 additional $1/\Gamma(t)$ (where $\Gamma(t)$ is the Lorentz factor
 of the blast at observer time t) beyond the angle into which the
 baryons have penetrated,
 covers their line of sight. Typical numbers might be a beam opening
 angle of $0.1$ radians, a penetration angle of 0.03 to 0.04
 radians, corresponding to $1/\Gamma_{fs}$ (where  $1/\Gamma_{fs}$
 is the Lorentz factor of the penetrating baryons (Levinson and
 Eichler 2003), and $1/\Gamma(t)$  for typical afterglow observation times
 of $\sim$10 hours is about 1/30 to 1/10. Although this is about
 enough to cover the entire range of viewing angles over
 which the GRB would be seen, it is just barely so.  Given the scatter in parameters
 so natural to astrophysical systems,  we might
 expect to see, every now and then, a baryon-underloaded GRB with
 little or very weak afterglow.
 Delayed-afterglow( Granot et al 2002) or
 briefly-delayed afterglow (Eichler 2005) is another possibility.

 In this letter we discuss whether  the observations of several
 dozen afterglows are consistent with the hypothesis that some
 afterglows have far less afterglow efficiency than the majority.
 We find that they are; there are four or five obvious
  outliers relative to an otherwise expected distribution of
  afterglow efficiencies clustering "near" unity (but see below).
  It cannot of course be proved that the
 reason for the poor afterglow efficiency is  baryon
 underloading. It may be due to a lower ambient density
 (e.g. Fan et al and references therein) that has
 the effect of
 spreading  the afterglow over a longer timescale  thus lowering
 the afterglow luminosity. However, this could be
 resolved with sufficiently thorough observations and a sufficiently
 large database.

    In order to minimize the likelihood of indirect correlations,
 we first recall that afterglow efficiency is correlated with
location of spectral peak $E_{peak}$ (Lloyd-Ronning and Zhang 2004
[LRZ]). Softer GRB seem to have lower blast efficiencies; the
efficiencies scale roughly as $E_{peak}^{\alpha}$ with
${1.4<\alpha<2}$ (see below). This, and its possible physical
interpretations, are discussed in section 2. We then plot the
afterglow efficiency corrected for this correlation against burst
duration and show a) that the data appear better organized after
the correction, b) that the majority of GRB have inferred blast
efficiencies of roughly 1/7, which could possibly be identified
with baryon saturation, given the uncertainties and c) that there
is no conclusive correlation with burst duration, with the present
sample. We also note that several GRB are outliers to this
correlation and all of them have anomalously {\it high} values for
$\epsilon_{\gamma}$.

 It is emphasized that the results are not meant to be convincing
beyond reasonable doubt. They are meant to show trends that we
suggest should be checked with the  much richer data set that
SWIFT should provide. The significance of the trends, if real,
would be some or all of the following implications: a) Most GRB
have blast energies that are at least somewhat
 lower than the $\gamma$-ray energies.  Previous
estimates may have been influenced by the preferential
underrepresentation of the $\gamma$-ray energies, relative to
afterglow energies, by off-beam observers.   b) While most GRB in
the data set cluster around a value of $\epsilon_{\gamma}$ of
order a few, several have extremely large values of
$\epsilon_{\gamma}$. These could plausibly be interpreted as
baryon-underloaded GRB. According to Freedman and Waxman (2001),
the blast energy estimate is independent of ambient density and
they cannot be interpreted as GRB that took place in an underdense
environment if the observed X-ray frequency is above the cooling
frequency, though it can be posited that the ambient density
and/or magnetic field energy was anomalously low and that the
cooling frequency was anomalously high.  c) There is some
indication that some of the anomalous GRB with very high
$\epsilon_{\gamma}$ tend to be short, and could thus be attributed
to a qualitatively different type of phenomenon and/or
environment. The search for afterglow from short GRB that can be
undertaken with SWIFT will thus be important. However, three of
the five lasted longer than 25 seconds and have no apparent
distinguishing characteristics other than a weak afterglow. d)
Various explanations for the Amati et al correlation can be tested
with a good enough data set.

\section{Afterglow Correlates with  $E_{peak}$.}

The values of $E_{peak}$ and  $E_{iso}$ correlate according to the
relation $E_{iso} \propto E_{peak}^2$ (Amati et al 2002, Atteia et
al 2004). Two possible accounts of the   $E_{iso} \propto
E_{peak}^2$ correlation are a) the dirty fireball model (e.g.
Dermer 1999, Qin et al 1998), in which baryon overloading delays
transparency until photons have softened to X-ray energies and b)
off-beam viewing, in which the observed $E_{peak}$ is lessened by
kinematic effects, viz. the reduced blue shift at the observer's
viewing angle relative to that seen by an observer in the beam
(EL04).

In the viewing angle model for the Amati et al relation proposed
by (EL04), the apparent total isotropic equivalent fluence is
viewer angle dependent. It is lowered by a viewing angle offset
from the closest part of the beam by angle $\theta$, approximately
as $D(\theta,\Gamma)^2$, where $D(\theta, \Gamma)\equiv
1/\Gamma(1-\beta cos\theta)$. This is opposed to the $D(\theta,
\Gamma)^3$ dependence that would apply to  a thin pencil beam
because  the
 solid angle that makes  a significant contribution to what is
detected by observers  just outside the beam is roughly
proportional to the  factor   $(1- cos\theta) \sim 1/(1-\beta
cos\theta)$. Earlier discussions of off-beam viewing (Yamazaki et
al 2002, 2004) predict a different $E_{peak}-E_{iso}$ relation and
attribute the Amati et al relation to an unspecified intrinsic
correlation. A second distinction to be noted in the context of
this paper is that here we are considering the possibility of
baryon poor lines of sight which nevertheless emit $\gamma$-rays.

The apparent afterglow fluence is also be reduced by off-beam
viewing, but generally not as much. Freedman and Waxman (2001)
noted that X-ray afterglow fluence  at $t \sim$10 hours could be
used as a calorimeter for the blast energy.  The Lorentz factor
after 10 hours, the typical time for BeppoSax measurements of
afterglow, $\Gamma_{x}$, is expected to be about a factor of 10
less (if the expansion is into a uniform medium) than the Lorentz
factor at 100 seconds, $\Gamma_p$. (The subscript p is for
"prompt", which refers to $t\le 100$s.) Hence the reduction in
prompt fluence {\it relative} to the fluence of X-ray afterglow is
given by
\begin{equation}
\frac{\epsilon_{\gamma}(\theta)}{\epsilon_{\gamma}(0)}
=[\left(D(\theta, \Gamma_p)
)/D(0,\Gamma_p)\right]^2/\left[D(\theta,\Gamma_x)/D(0,\Gamma_x)\right]^2
\label{effa}
\end{equation}

By hypothesis that $E_{peak}$ is established by viewing angle
effects,

\begin{equation}
E_{peak}(\theta)/E_{peak}(0) = D(\theta, \Gamma_p)/D(0,\Gamma_p)
\end{equation}

After using (2) to eliminate the viewing angle in favor in
$E_{peak}$, equation (1) becomes:
\begin{equation}
\frac{\epsilon_{\gamma}(\theta)}{\epsilon_{\gamma}(0)}=
\frac{E_{peak}(\theta)}{E_{peak}(0)}
\left[\frac{(1-\frac{\beta_x}{\beta_P})+(\frac{\beta_x}{\beta_P}-\beta_x)
(\frac{E_{peak}(\theta)}{E_{peak}(0)})^{-\frac{1}{2}}}{(1-\beta_x)}\right]^2
\end{equation}

Over the range of viewing angles $\theta \Gamma_x \ll 1$,
\begin{equation}
\frac{\epsilon_{\gamma}(\theta)}{\epsilon_{\gamma}(0)}\simeq
\left[\frac{E_{peak}(\theta)}{E_{peak}(0)}\right]^2
\end{equation}

Thus the viewing angle explanation for the Amati et al relation
predicts that the apparent ratio of gamma ray energy to blast
energy $\epsilon_{\gamma} \equiv E_{\gamma,iso}/E_{k,iso}$ should
decrease as $E_{peak}^2$ decreases, as described by equation (4).
Weak correlation in the intrinsic $E_{peak}$ with the opening
angle (see Fig. 6 below) and the fact that the beam probably does
not have a sharp edge could cause the correlation to deviate
somewhat from equation (4). So might other indirect correlations.
Also, pole to equator energy transfer, a true physical effect
(Kumar \& Granot 2003), may play some role. In any event, we
expect the qualitative correlation to survive these
considerations.

\section{Radiative Efficiency and Spectral Peak}
We have plotted $\gamma$-ray efficiency $\epsilon_{\gamma}$
against $E_{peak}^2$, defining efficiency as $\frac{E_{\gamma
,iso}}{E_{k}}$ where $E_k$ is the isotropic equivalent kinetic
energy of the GRB ejecta. We have used the $E_k$ values as
presented in LRZ based on X-ray afterglow luminosity in Berger et
al (2003) We add more data to the plot in Figure 7 of LRZ, using
the correlation between X-ray afterglow at ten hours and blast
energy to estimate $E_k$ for GRB 980326, GRB 980329 and GRB
000214.

The efficiency GRB 990506 has been taken from Freedman \& Waxman
(2001). The data for GRB 980329 have a large uncertainty as the
redshift is uncertain ($2<z<3.9$). We have used the estimates by
Ghirlanda et al for $ E_{peak}(1 + z)$ and $E_{\gamma ,iso}$.
Using equation (1) in Berger et al (2003), and the relation
between $L_{x,45}(10\;hr)$ and $E_{k,52}$ of LRZ,\footnote{In this
paper we use the usual cgs subscript convention: $Q = 10^{x}Q_x$.}
this would give GRB 980329 an isotropic blast kinetic energy value
of $E_{k,52} = 2.1\pm1.0$.

The radiative efficiency we calculate differs from that as
calculated by Freedman \& Waxman (2001) who used an assumed value
for the redshift for their calculations. For all GRBs, we have
used values for $E_{peak}(1+z),\;T_{90}$ and $E_{\gamma ,iso}$
from Ghirlanda et al (2004), Bloom et al (2003), and the
\emph{HETE} webpage. We also add GRB 020124 whose blast energy we
estimate below. Finally, the discovery of the host galaxy of GRB
040924 has allowed the redshift to be measured at $z = 0.859$
(Starling et al: GCN GRB Observation Report) as well as $E_{\gamma
,iso}$. We have taken $E_{peak}$ from the \emph{HETE} webpage.
Below we estimate the kinetic energy of the GRB 040924.
\\ \\

\emph{GRB 040924:} The afterglow of GRB 040924 is reproduced in
Fan et al (2004), and can be extrapolated for
$F_{\nu,max}\geq250\;\mu Jy$, using
$F_\nu\;=\;F_{max}(\frac{\nu}{\nu_m})^{\frac{-(p-1)}{2}}$ (Sari et
al., 1999). In this case $p=2.42$, and $t_{d} = 1.09\times10^{-2}$
(Fan et al 2004). The $E_k$ and its uncertainty have been
calculated using the equations of adiabatic afterglow evolution
(Sari et al., 1999) as arranged below, with $D_{28} = 1.68$:

\begin{equation}
      E_{k,52}\; = (4.03\times10^{-36})\;\nu_{m}^{2}\;\epsilon_{e}^{-4}\;\epsilon_{B}^{-1}
\end{equation}
\begin{equation}
      E_{k,52}\; =
      (2.56\times10^{-5})\;F_{\nu ,max}\;\epsilon_{B}^{-\frac{1}{2}}\;n_0^{-\frac{1}{2}}
\end{equation}
where $\nu_m$ is in hz and $F_{\nu, max}$ is in $\mu$Jy. This
allows upper and lower bounds to $E_k$ to be calculated, with
reasonable ranges assumed for the unknown parameters:
$\epsilon_{e}\;[0.03,0.3]$,\; $\epsilon_{B}\; [10^{-3},10^{-2}]$
and $n_0\; [0.01,3] (cm^{-3})$.\\ \\These equations are plotted
below in Fig.1 to show the bounds imposed on $E_k$ as a function
of $F_{\nu ,max}$.  We include in this plot, the lines
corresponding to equation (6) above for the where the circumburst
density $n_{0}\;=\;1 $.

\begin{figure}
\includegraphics[height = 2.8 in]{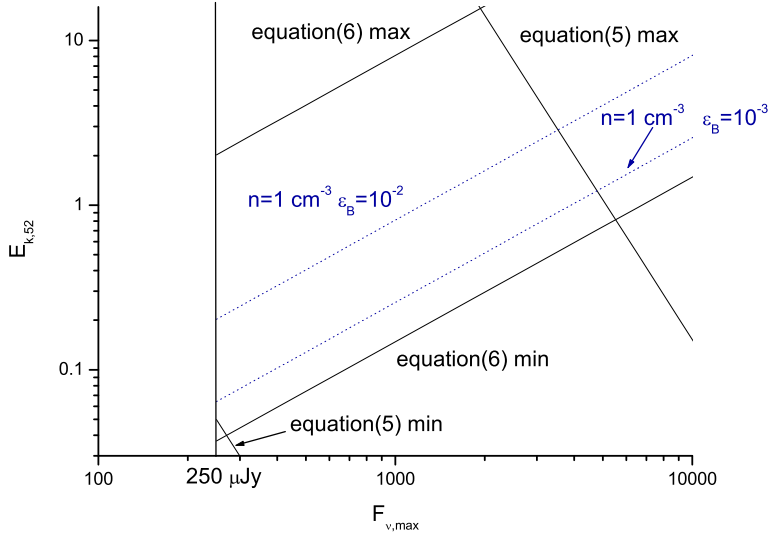}
\caption{Bounds imposed on $E_{k}$ from the two equations above,
for all $F_{\nu,max}\geq250 \mu Jy$. The line 'Equation (6)min' is
a limit imposed by equation (6) assuming $n\;=\;3\; cm^{-3}$ and
$\epsilon_{B}\;=\; 0.01$. $E_{k}$ is excluded below this line.
Lines '(6) max', '(5) min' and '(5) max' show the other relevant
limits as defined in the text. The dotted lines correspond to
limits where $n = 1\;cm^{-3}$ for comparison.} \label{Fig. 1}
\end{figure}
\begin{figure}
\includegraphics[height = 3.0 in]{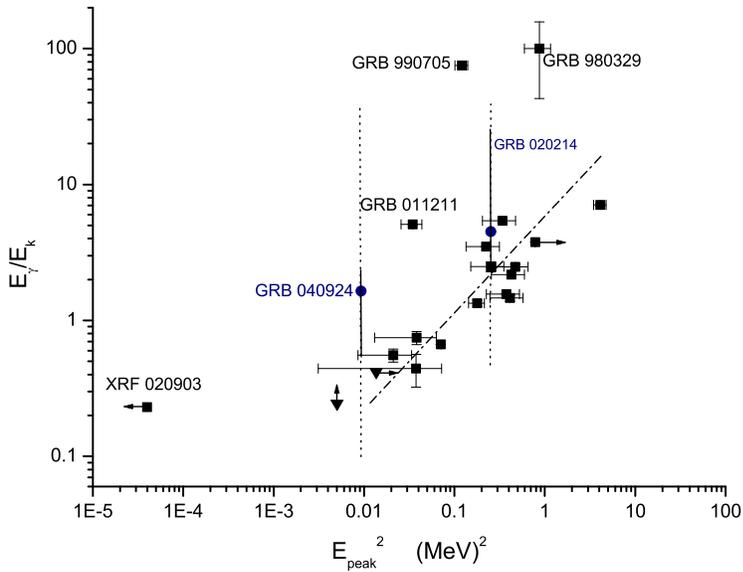}
\caption{Radiative efficiency and spectral peak. The squares are
data from LRZ, and the triangles correspond to GRBs 980326 and
000214. The circles correspond to the most likely efficiencies for
GRB 040924 and GRB 020124, with dotted lines extending to the
extremes and solid lines covering the total uncertainty for the
case where the circumburst density $n=1\;cm^{-3}$. The
dashed-dotted line is described in the text.} \label{Fig. 2}
\end{figure}

The most likely value for $E_{k,52}$ has been taken as the center
of the polygon bounded by the equations and the line $F_{\nu ,max}
= 250 \mu Jy$. We add the extreme cases as well as the limits on
$E_{k}$ that result from assuming $n = 1\;cm^{-3}$, for comparison
in our results. We find the best value of $E_{k,52}$ to be $0.9$
with extremes at $0.04 < E_{k,52} < 15$ and a range for the case
$n_0 = 1$ of
\;$0.65<E_{k,52}<2.75$.\\ \\
\emph{GRB 020124:}By the same method employed for GRB 040924, we
have used the afterglow lightcurve produced in Berger et al (2002)
to calculate the kinetic energy of GRB 020124 over the same ranges
for the unknown parameters, using $D_{28} = 8.38$. We found that
in the limits of uncertainty $0.74 < E_{k,52} < 70.5$, and for the
case $n = 1\;cm^{-3}$, $1.22 < E_{k,52} < 11.61$, and the most
likely blast energy to be: $E_{k,52}\simeq 4.51$.

\section{Results}
In Fig.2, the radiative efficiency of each GRB is plotted as a
function of the square of the spectral peak. There are 17 GRBs
closely correlated along the dashed-dotted line of  best fit, with
the remaining 5 outliers being XRF 020903 (eight standard
deviations $(8.0\sigma)$ above the line), our estimate for GRB
040924 $(4.2\sigma)$, GRB 011211 $(4.5\sigma)$, GRB 990705
$(7.9\sigma)$, and GRB 980329 $(5.7\sigma)$, where the standard
deviation $\sigma$ is computed for log$\frac{E_{\gamma}}{E_k}$
relative to the corresponding value given by the linear fit;
$\sigma = 0.23$ for the group of 17 well correlated GRBs. The
existence of these bursts with exceptional radiative efficiency,
roughly an order of magnitude above the majority for a given
$E_{peak}$, provides evidence of a distinct subclass of GRBs. The
slope of the line of best fit in this plane reveals a correlation
whereby $\frac{E_{\gamma}}{E_{k}}\;\sim\; E_{peak}^{1.5}$.
\\ \\

\noindent\emph{Radiative Efficiency, Spectral Peak and $T_{90}$:}
We include a graph (fig.3) of $\gamma$-ray efficiency corrected
for the correlation noted above, as a function of $T_{90}$, taking
$T_{90}$ from Ghirlanda et al (2004) or the \emph{HETE} webpage.

\begin{figure}
\includegraphics[height = 3.0 in]{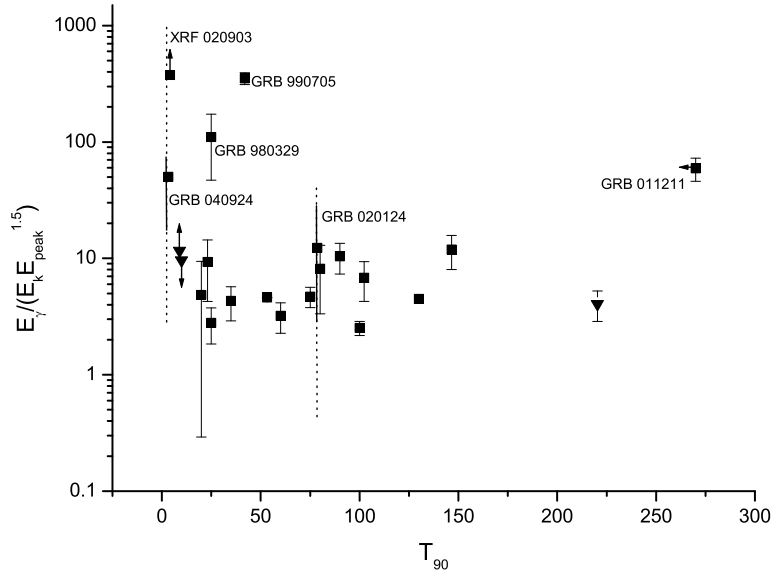}
\caption{Radiative efficiency over spectral peak, corrected for
the $E_{peak}^{1.5}$ correlation, plotted against $T_{90}$. The
triangles are GRBs 980326, 990506 and 000214. For the uncertain
cases of GRB 040924 and GRB 020124, the dotted lines extend to the
possible extremes, and the solid lines cover the range for the
case where $n=1.0\;cm^{-3}$.} \label{Fig. 3}
\end{figure}
\begin{figure}[htb]
\includegraphics[height = 3.0 in]{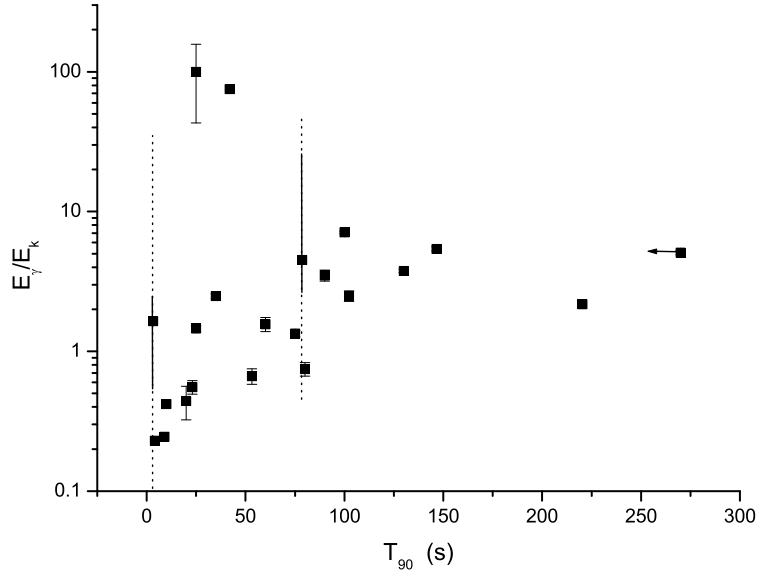}
\caption{Radiative efficiency, uncorrected for the $E_{peak}$
correlation, plotted against $T_{90}$.} \label{Fig. 4}
\end{figure}

In this plane, there appears to be a general population of well
correlated GRBs and several scattered outliers. The majority of
bursts appear to settle on a roughly constant value
$\frac{E_{\gamma}}{E_{k}E_{peak}^{1.5}} \simeq 7\; (MeV)^{-1.5}$,
with a scatter of less than one order of magnitude. Contrast this
with Fig 4., where the naive efficiency (i.e. uncorrected for
$E_{peak}^{1.5}$) has either a larger scatter or some dependence
on $T_{90}$. The outliers XRF 020903 and GRB 040924 are rather
short bursts, though still at least $\sim 1$s. However, GRB
990705, GRB 980329 and GRB 011211 all lasted at least 25s.\\

\noindent\emph{Viewing angle calculations and radiative
efficiency:} If we can assume that viewing angle on the jet is the
\emph{only} factor which reduces an otherwise standard radiative
efficiency and standard spectral peak of GRBs, then we can compare
equation (3) with the plot we have made in Fig. 2, normalising
both the efficiency and $E_{peak}$ to be unity for observers where
$\theta = 0$.
\begin{figure}
\includegraphics[height = 3.0 in]{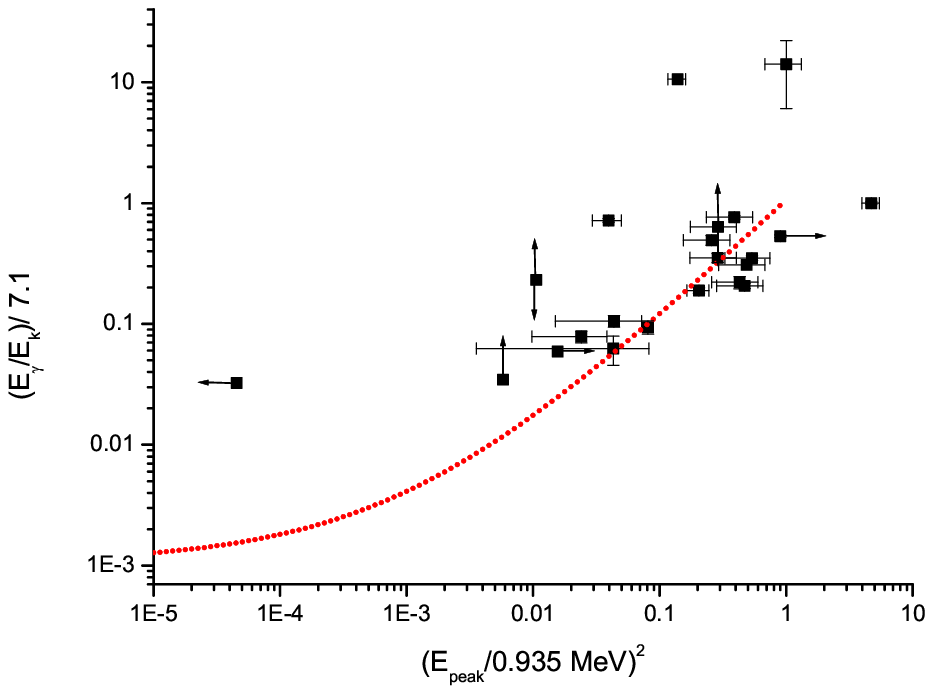}
\caption{Radiative efficiency and spectral peak compared with a
model for viewing angle reduction.}
\label{Fig. 5}
\end{figure}
\begin{figure}
\includegraphics[height = 3.0 in]{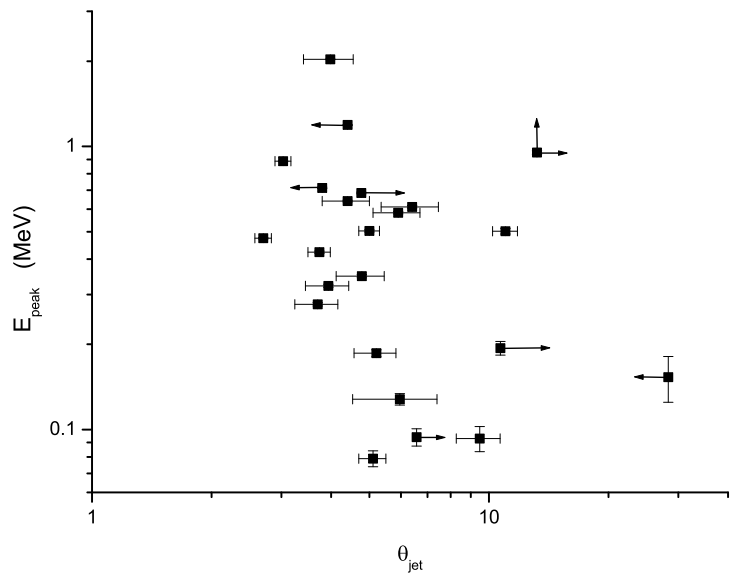}
\caption{Spectral peak and jet opening angle. The data are from
Ghirlanda et al (2004). Although the correlation is not strong,
there is nonetheless a slight indication that wider jets have
lower spectral peaks.} \label{Fig. 6}
\end{figure}

\section{Discussion}
The question of whether short bursts have afterglows is a
longstanding one. If they result from neutron star coalescence
(Goodman 1986, Paczynski 1986, Eichler et al 1989, Paczynski
1990), then they might take place in regions of low ambient
density (Fan et al 2004 and references therein), which would
weaken and prolong  their afterglow. Here we have called attention
to several weak afterglows whose GRBs were not so short, such as
GRBs 011211, 990705, and 980329. We know of no \emph{a priori}
particular reason for them to have had weak afterglows and suggest
that they may have been the occasional bursts that we view along
baryon-poor lines of sight.

With mounting evidence that GRBs may be divided into subclasses
which shed light on jet structure or the cause of GRBs (see Bloom
et al 2003), our hypothesis would add yet another  distinction
between populations of GRBs. Relying on previous work, we find
that for the GRBs with available data, 5 of 22 appear unassociated
with what is otherwise a closely clustered population in blast
efficiencies. All of them have a weak afterglow, (high radiative
efficiency), whereas none were particularly deviant in the
opposite direction. We have speculated on possible trends in this
small sample, but stress our anticipation of future data.

The majority of burst efficiencies follow $\frac{E_{\gamma}}{E_k}
\sim E_{peak}^{1.5}$. Ghirlanda et al (2004) find a correlation
between collimation corrected $E_{\gamma}$ and $E_{peak}$ whereby
$E_{\gamma}\sim E_{peak}^{1.4}$. (The slightly shallower than
$E_{peak}^2$ dependence found by Ghirlanda et al. could be
attributed at least in part to a removal in the intrinsic scatter
of $E_{iso}$, if wider beams tend to have lower $E_{peak}$. From
Fig. 6, it appears as though they do but we do not yet know the
reason. Perhaps a wider beam allows more opportunity for off-beam
viewing.) In any case, there may be additional causes for scatter
in $E_{peak}$ such as a dirty fireball effect, which would lower
both $E_{peak}$ and $E_{\gamma}$ but not $E_k$. However, it would
do so at the cost of imposing an extremely small radiative
efficiency. Curiously, we find that $\gamma$-ray efficiency has
nearly the same dependence on $E_{peak}$ as does $E_{\gamma}$ in
the correlation noted by Ghirlanda et al. This could be attributed
to the simple fact that apparent $\gamma$-ray luminosity depends
more on viewing angle than does apparent afterglow luminosity.

It is anticipated that SWIFT data will reveal whether there are
more GRBs distinct from the majority by their high radiative
efficiency.  The sample we have used is not free of all bias; on
the other hand, the sample of available redshifts may itself
suffer a possible selection bias against weak afterglows, so the
eventual fraction of weak afterglow GRB may be considerably
different from the 5 of 22 portrayed here.

The significance of this result, if valid,  is that the blast
energy as a fraction of the total is only about 1/7, and that
instances where it is greater can be largely attributed to viewing
angle dependent reduction of the apparent radiative efficiency.
Theoretical estimates for the dissipation efficiency of the
internal shocks vary (Kumar 1999, Guetta, Spada, Waxman, 2001,
Beloborodov, 2000, Kobayashi, Sari, 2001), but in principle this
efficiency can be large. Given the uncertainties and possible
systematic errors, both in afterglow observations and in the
theory, the value of 7 for the ratio of isotropic equivalent
$\gamma$-ray energy to isotropic equivalent kinetic energy could
be interpreted as a not implausible value for a baryon-saturated
outflow. However, it may be uncomfortably large for the scenario
in which internal shocks in a baryonic
outflow convert kinetic energy to $\gamma$-ray energy.\\

The authors acknowledge a Center of Excellence grant from the
Israel Science Foundation and the grant from from the Israel-U.S.
Binational Science Foundation and support from the Arnow Chair of
Theoretical Astrophysics.



\begin{thebibliography}{99}
\bibitem[]{1} Amati, L., et al 2002 A \& A, 390, 81
\bibitem[]{2} Atteia, J-L., et al 2004 AIPC, 727, 37A
\bibitem[]{3} Beloborodov, A.M. 2000, ApJ, 539, L25
\bibitem[]{4} Berger, E., Kulkarni, S.R., and Frail, D.A., 2003, ApJ.
590, 379
\bibitem[]{5} Berger, E., et al, 2002, ApJ. 581, 981
\bibitem[]{6} Bloom, J.S., Frail, D.A., and Kulkarni, S.R., 2003,
ApJ. 594, 674
\bibitem[]{7} Dermer, C. 1999 \emph{Astron. Astrophys. Suppl Ser}, 138, 519
\bibitem[]{8} Eichler, D., et al 1989, Nature, 340, 126
\bibitem[]{9} Eichler, D., \& Levinson, A. 1999 ApJ, 521, L117
\bibitem[]{10} Eichler, D. \& Levinson, A., 2003 ApJ, 594, L119
\bibitem[]{11} Eichler, D. \& Levinson, A., 2004 ApJ, 614, L13
\bibitem[]{12} Fan, Y.Z., Zhang, B., Kobayashi, S., and
Meszaros, P., 2004 American Astronomical Society Meeting 205,
160.07 \
\bibitem[]{13} Freedman, D.L., \& Waxman, E., 2001 ApJ, 547, 922
\bibitem[]{14} Ghirlanda, G., Ghisellini, G., and Lazzati, D., 2002, ApJ, 616,
331
\bibitem[]{15} Goodman, J., 1986, ApJ, 308, L47
\bibitem[]{} Granot, J., Panaitescu, A., Kumar, P. and Woosley, S.
(2002), ApJ, 570, L61
\bibitem[]{16} Guetta, D., Spada, M., \& Waxman, E. 2001, ApJ, 557, 399
\bibitem[]{17} Katz, J., \& Canel, L., 1996, ApJ, 471, 915
\bibitem[]{18} Kobayashi, S., \& Sari,R. 2001, ApJ, 551, 934
\bibitem[]{19} Kumar, P., 1999, ApJ, 523, L113
\bibitem[]{20} Kumar, P., \& Granot, J., 2003, ApJ, 591, 1075
\bibitem[]{21} Lloyd-Ronning, N., and Zhang., B, 2004, ApJ, 613, 477
\bibitem[]{22} Lyutikov, M., \& Blandford, R., 2004 astro-ph/0312347
\bibitem[]{23} Meszaros, P., \& Rees, M., 1992, ApJ, 397, 570
\bibitem[]{24} Meszaros \& Rees 1994 \emph{Mon Not Astron Soc}, 269, L41
\bibitem[]{25} Narayan, R., Paczynski, D., \& Piran, T., 1992, ApJ,
395, L83
\bibitem[]{26} Paczynski, B., 1986, ApJ (Letters), 308, L43
\bibitem[]{27} Paczynski, B., 1990, ApJ, 348, 485
\bibitem[]{28} Qin, B. et al 1998 ApJ 494, L57
\bibitem[]{29} Rees \& Meszaros 1992 \emph{Mon. Not R. astr Soc}, 258, 41P
\bibitem[]{30} Sari, R., Piran, T., and Narayan, R., 1998, ApJ, 497,
L17
\bibitem[]{27} Yamazaki, R., Ioka, K., \& Nakamura, T. 2004, ApJ, 606, L33
\bibitem[]{28} Yamazaki, R., Ioka, K., \& Nakamura, T. 2002, ApJ, 571, L31
\bibitem[]{31} Zand, J., et al, 2000, ApJ, 545, 266
\bibitem[]{32} http://space.mit.edu/HETE/Bursts/GRB040924/
\bibitem[]{33} Frontera, F. et al, GCN GRB OBSERVATION REPORT 1215, http://gcn.gsfc.nasa.gov/other/011211.gcn3
\bibitem[]{34} Starling, R. et al, GCN GRB OBSERVATION REPORT 2800, http://gcn.gsfc.nasa.gov/gcn/other/040924.gcn3
\end{thebibliography}
\end{document}